\documentclass[nocite]{epl}

\title{High-resolution hyperfine spectroscopy of excited
states using electromagnetically-induced transparency}
\shorttitle{Hyperfine spectroscopy of excited states}
\author{Anusha Krishna, Kanhaiya Pandey, Ajay Wasan, and Vasant Natarajan
\thanks{E-mail: \email{vasant@physics.iisc.ernet.in}}}
\shortauthor{Anusha Krishna \etal}
\institute{Department of Physics, Indian Institute of
Science, Bangalore 560 012,  INDIA}

\begin{document}
\bibliographystyle{prsty}
\maketitle

\begin{abstract}
We use the phenomenon of electromagnetically-induced
transparency in a three-level atomic system for hyperfine
spectroscopy of upper states that are not directly coupled
to the ground state. The three levels form a ladder system:
the probe laser couples the ground state to the lower
excited state, while the control laser couples the two
upper states. As the frequency of the control laser is
scanned, the probe absorption shows transparency peaks
whenever the control laser is resonant with a hyperfine
level of the upper state. As an illustration of the
technique, we measure hyperfine structure in the $7S_{1/2}$
states of $^{85}$Rb and $^{87}$Rb, and obtain an
improvement of more than an order of magnitude over
previous values.
\end{abstract}


The use of coherent-control techniques in three-level
systems is now an important tool for modifying the
absorption properties of a weak probe laser
\cite{NSO90,VAR96,RWN03}. For example, in the phenomenon of
electromagnetically induced transparency (EIT), an
initially absorbing medium is made transparent to a probe
beam when a strong control laser is switched on
\cite{BIH91,HAR97}. EIT techniques have several practical
applications in probe amplification \cite{MEA00}, lasing
without inversion \cite{ZLN95}, and suppression of
spontaneous emission \cite{RWN03,GZM91,ZNS95,ZSC96}.
Experimental observations of EIT have been mainly done
using alkali atoms (such as Rb and Cs), where the
transitions have strong oscillator strengths and can be
accessed with low-cost tunable diode lasers.

In this paper, we use the phenomenon of EIT in a novel
application, namely high-resolution spectroscopy of
hyperfine structure in excited states. The experiments are
done in a ladder system, where the control laser drives the
upper transition and the probe laser measures absorption on
the lower transition. In normal EIT experiments, the
frequency of the probe laser is scanned while the frequency
of the control laser is kept fixed. By contrast, in our
technique, it is the frequency of the control laser that is
scanned while the probe laser remains locked on resonance.
The probe-absorption signal then shows transparency peaks
every time the control laser comes into resonance with a
hyperfine level of the excited state.

Measurement of hyperfine structure in excited states is
important because these states are used in diverse
experiments ranging from atomic signatures of parity
non-conservation (PNC) to resonance ionization mass
spectrometry \cite{PDT94}. For example, the $7S$ state in
Cs has been used for the most sensitive test of atomic PNC
to date \cite{WBC97}. However, to convert the experimental
results into useful information about the parity-violating
weak interaction, the data has to be compared to complex
theoretical calculations in Cs. Knowledge of hyperfine
structure thus provides valuable information about the
structure of the nucleus (nuclear deformation) and its
influence on the atomic wavefunction. Hyperfine
spectroscopy on these excited states is complicated by the
fact that they have the same parity as the ground state and
can only be accessed through weak two-photon transitions.
By contrast, our technique provides strong, easily-measured
signals. In addition, there is an interesting narrowing of
the linewidth in room-temperature vapor by the use of
counter-propagating beams, which is important for high
accuracy in the measurement. As an illustration of the
power of this technique, we have used it to measure
hyperfine structure in the $7S_{1/2}$ state of Rb, and
demonstrate an improvement of more than an order of
magnitude in precision over previous values.

\begin{figure}
\twofigures[height=4.5cm]{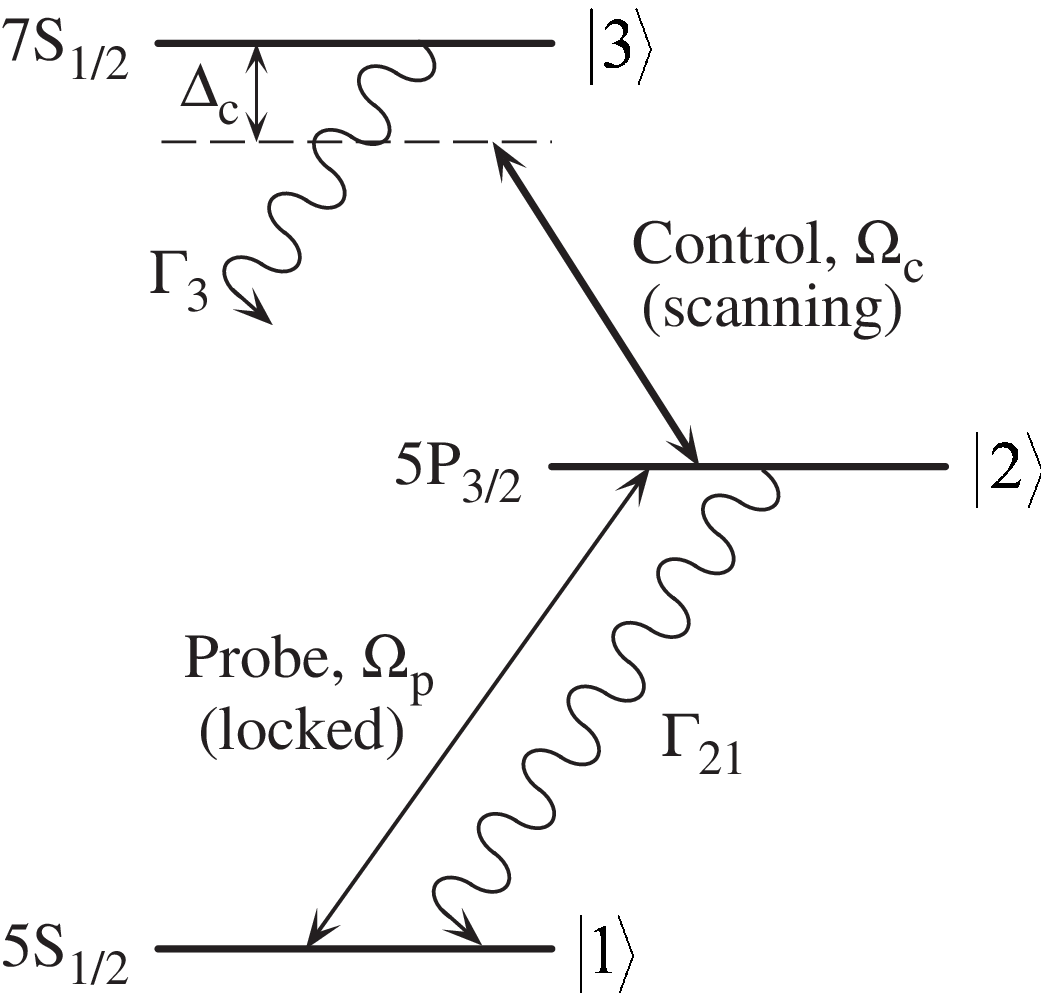}{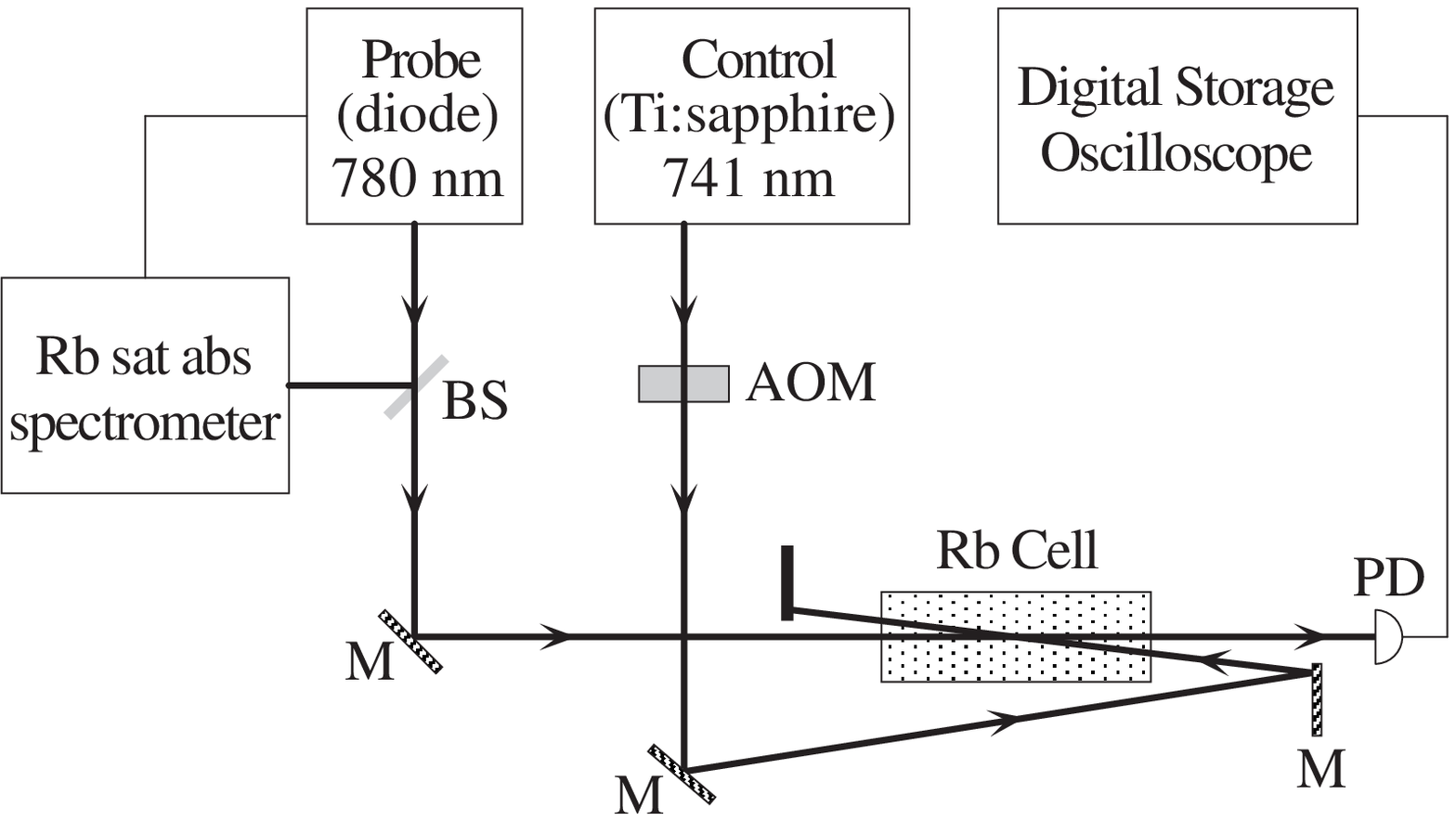}
\caption{Three-level ladder system in Rb. The probe laser
has Rabi frequency $\Omega_p$ and is locked to the lower
transition. The control laser has Rabi frequency $\Omega_c$
and is scanned across the upper transition with a detuning
$\Delta_c$.} \label{f1}

\caption{Schematic of the experiment. The control beam
after the AOM consists of two parts, one at the laser
frequency and one shifted by the rf frequency. The angle
between the beams in the vapor cell has been exaggerated
for clarity, in reality it is less than 10 mrad. Figure
key: Sat.\ abs., saturated absorption; BS, beam splitter;
M, mirror; AOM, acousto-optic modulator; PD, photodiode.}
\label{f2}
\end{figure}

The ladder system in Rb is shown in Fig.\ \ref{f1}. The
levels $|1 \rangle$, $|2 \rangle$, and $|3 \rangle$, are
the $5S_{1/2}$, $5P_{3/2}$, and $7S_{1/2}$ states,
respectively. The probe laser is tuned to the lower
$5S_{1/2} \leftrightarrow 5P_{3/2}$ transition at 780 nm
with a Rabi frequency of $\Omega_p$. The spontaneous decay
rate from this state ($\Gamma_{21}$) is 6 MHz. The control
laser is tuned to the upper $5P_{3/2} \leftrightarrow
7S_{1/2}$ transition at 741 nm with a detuning of
$\Delta_c$ and Rabi frequency of $\Omega_c$. The
spontaneous decay from this state back to the ground state
is primarily through the cascade $7S \rightarrow 6P
\rightarrow 5S$ transition. The decay rate $\Gamma_3$ is 11
MHz.

The experimental set up is shown schematically in Fig.\
\ref{f2}. The probe beam is derived from a
feedback-stabilized diode laser system operating at 780 nm
\cite{BRW01}. The linewidth of the laser after
stabilization is less than 500 kHz. The probe beam has
$1/e^2$ diameter of 2 mm. The control beam comes from a
ring-cavity Ti:Sapphire laser tuned to 741 nm. The Ti:S
laser is stabilized to an ovenized reference cavity that
gives it an instantaneous linewidth of 500 kHz. The control
beam consists of two parts, one at the laser frequency and
one that is frequency shifted using an acousto-optic
modulator (AOM). As discussed later, the AOM-shifted beam
is used for calibrating the frequency-scan axis. One
important experimental requirement is to double pass
through the AOM so that the unshifted (zeroth-order) and
shifted beams propagate in the same direction. This ensures
that the two components are mixed perfectly and there is no
angle between them. The control beam is similar to the
probe beam and has a diameter of 2 mm. The beam
counter-propagates with respect to the probe beam through a
room-temperature vapor cell containing Rb. The absorption
through the cell is about 25\%. The two beams have
identical polarization and the angle between them is less
than 10 mrad.

\begin{figure}
\twoimages[height=5cm]{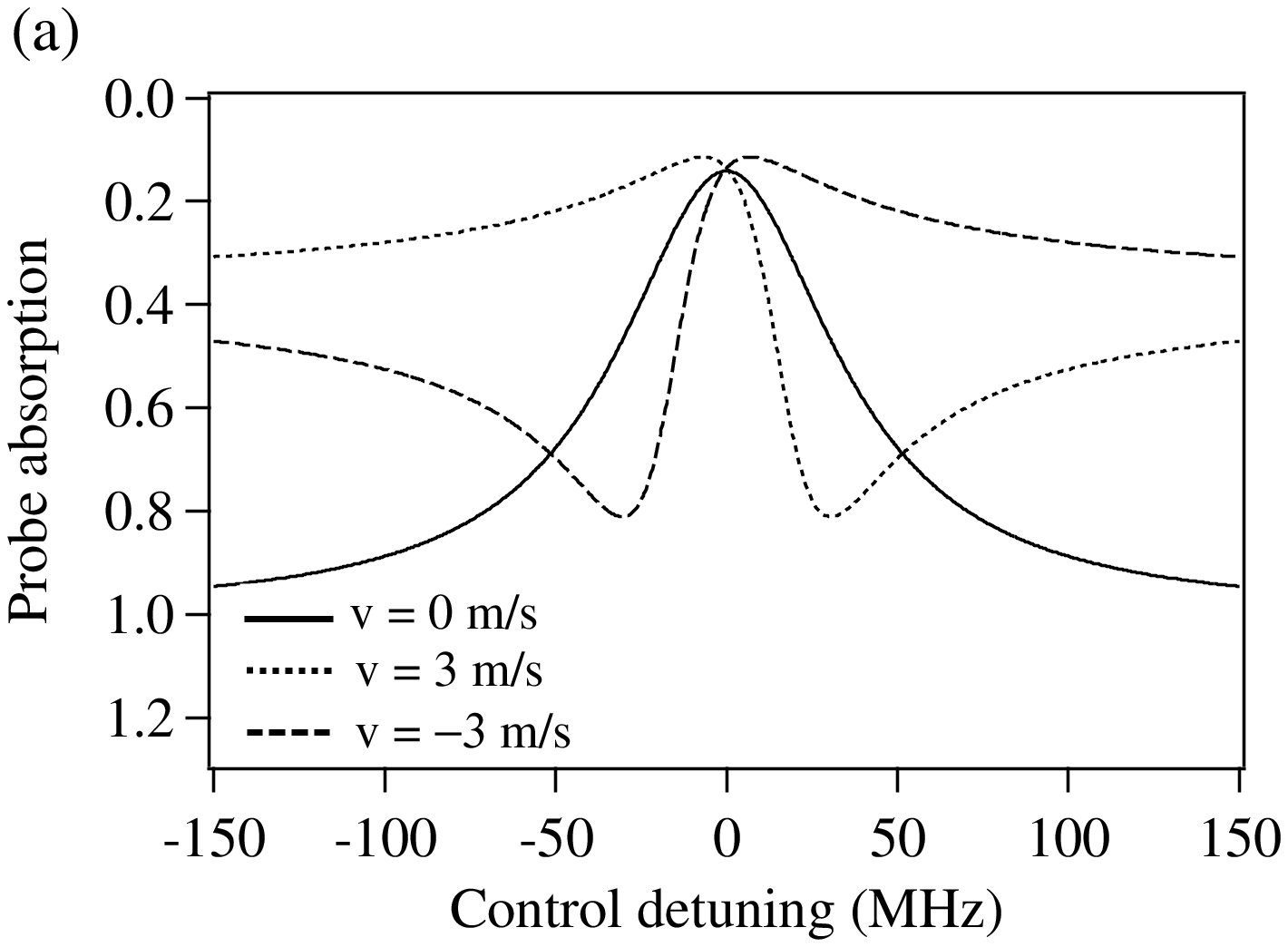}{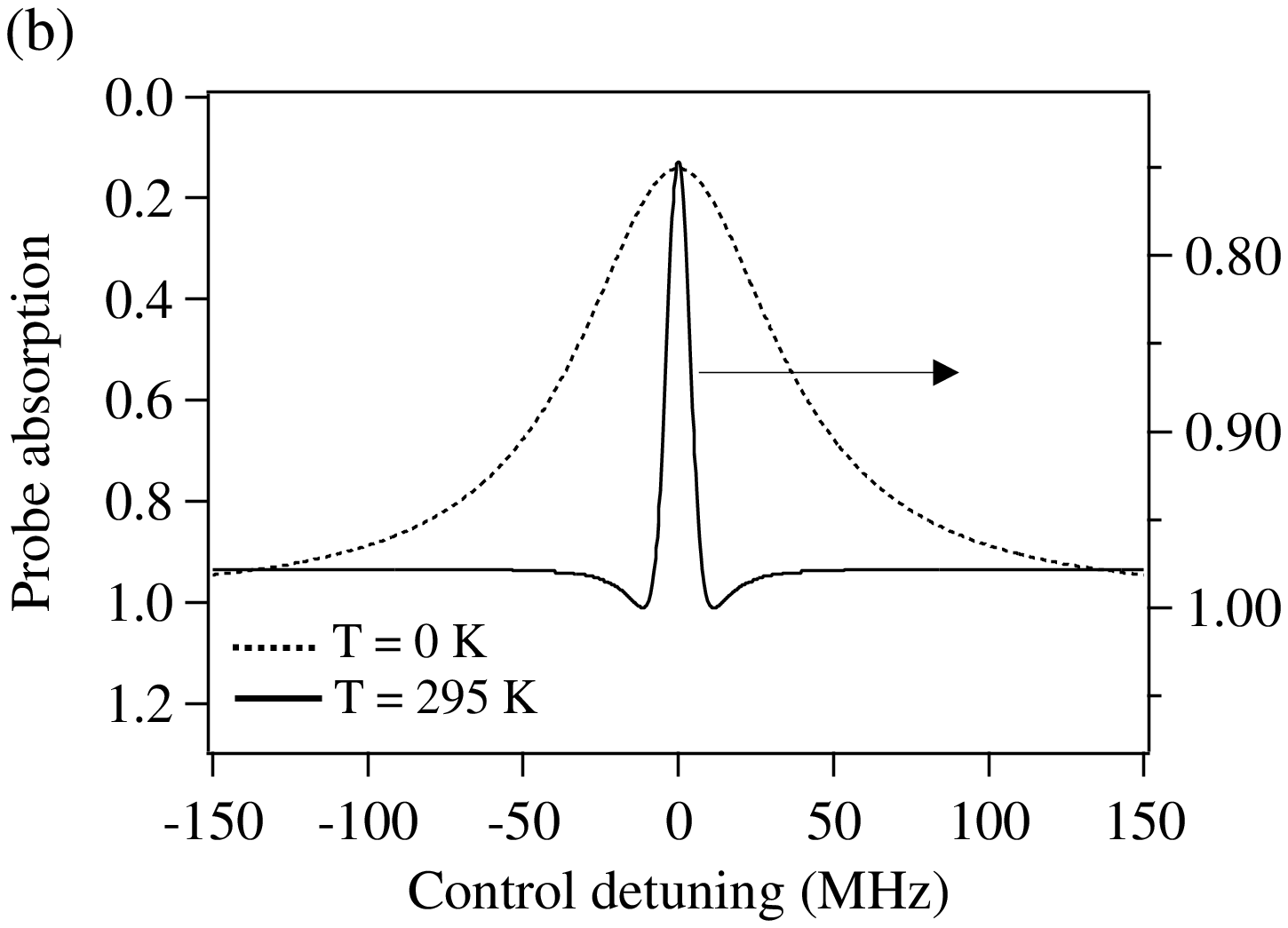}
\caption{Theoretical lineshapes for probe absorption. In
(a), we show the normalized absorption curves with
$\Delta_p = 0$ and $\Omega_c=20$ MHz, for zero-velocity
atoms and atoms moving with $v=\pm 3$ m/s. In (b), we show
the effect of thermal averaging: the linewidth is reduced
significantly at $T=295$ K compared to $T=0$ K. The scale
of the transparency is also reduced, as seen from the right
axis for the $T=295$ K curve. Note the inverted $y$-axes in
both graphs.} \label{f3}
\end{figure}

We first consider the theoretical analysis of probe
absorption in a ladder system in the presence of the
control laser. In the weak probe limit, the absorption of
the probe is proportional to Im$(\rho_{21})$, where
$\rho_{21}$ is the induced polarization on the $\left| 1
\right> \leftrightarrow \left| 2 \right>$ transition
coupled by the laser. From the density-matrix equations,
the steady-state value of $\rho_{21}$ is given by
\cite{GLJ95}:
\begin{equation}
\rho_{21} = - \frac{i(\Omega_p/2)}{ \left(
\frac{\Gamma_{21}}{2} - i \Delta_p \right) +
\frac{\Omega_c^2/4}{\Gamma_3/2 - i(\Delta_p + \Delta_c)}}
\end{equation}
where $\Delta_p$ is the detuning of the probe from
resonance. The calculated absorption curve (note the
inverted $y$-axis) for $\Delta_p =0$ and $\Omega_c=20$ MHz
is shown in Fig.\ \ref{f3}(a) as a function of $\Delta_c$.
The absorption is normalized to a value of 1 in the absence
of the control laser. There is a distinct EIT peak at line
center: the absorption falls by about 80\% when the control
laser is exactly on resonance. However, the linewidth of
the transparency peak is quite large, with a value of about
70 MHz for $\Omega_c=20$ MHz.

The above analysis is correct for a stationary atom. In a
room-temperature vapor, we have to account for the thermal
velocity distribution. For an atom moving with a velocity
$v$ in the same direction as the probe beam, the probe
detuning decreases to $\Delta_p - v/\lambda_p$, while the
control detuning increases to $\Delta_c + v/\lambda_c$ due
to the counter-propagating configuration. Here,
$\lambda_{p,c}$ are the respective center wavelengths for
the two transitions. The resulting absorption curves for
$v=3$ m/s and $v=-3$ m/s are also shown in Fig.\
\ref{f3}(a). The curve has a dispersive lineshape, and
shows increased absorption on one side of line center. More
importantly, it is slightly asymmetric because of the
unequal values of $\lambda_p$ and $\lambda_c$, i.e., for a
given velocity, the detuning changes by different amounts
for the probe and the control.

The effect of this for a Maxwell-Boltzmann velocity
distribution is seen in Fig.\ \ref{f3}(b). The two
absorption curves shown are for $T=0$ K and $T=295$ K. The
interesting feature is that the linewidth of the curve for
room-temperature atoms is significantly {\it reduced} by
thermal averaging; the use of a counter-propagating
geometry {\it decreases} the linewidth from 70 MHz for
zero-temperature atoms to 17 MHz for room-temperature
atoms. The narrowing is advantageous because it allows us
to measure hyperfine intervals with greater precision.
However, the narrowing is accompanied by a decrease in the
scale of the transparency, with only 25\% reduction in
absorption at line center. Note also the slight lineshape
distortion for room-temperature atoms, where the absorption
increases slightly before approaching an asymptotic value.
The cause for this is again the asymmetric detuning of the
control and probe for a given velocity. The distortion
disappears when the value of $\Omega_c$ increases beyond
100 MHz. Higher values of $\Omega_c$ also result in power
broadening of the linewidth, to about 50 MHz for
$\Omega_c=100$ MHz. Finally, because we are considering a
fixed frequency for the probe beam, the spectrum has a flat
background (corresponding to absorption by the
zero-velocity atoms) and changes only when the control beam
is close to resonance. This is different from other
high-resolution spectroscopy techniques in room-temperature
vapor, such as saturated-absorption spectroscopy, where the
spectrum shows an underlying Doppler profile.

\begin{figure}
\twoimages[width=7cm]{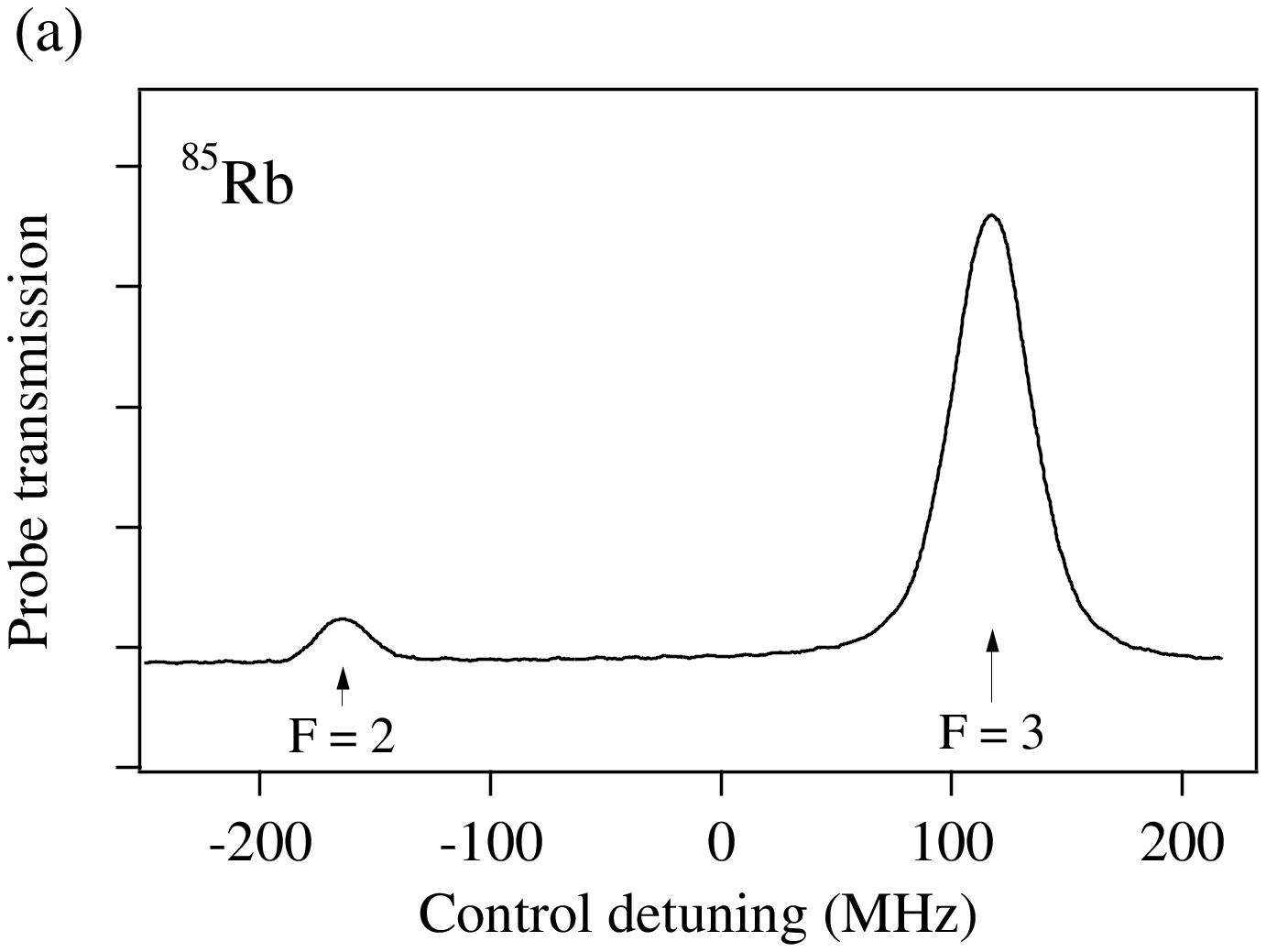}{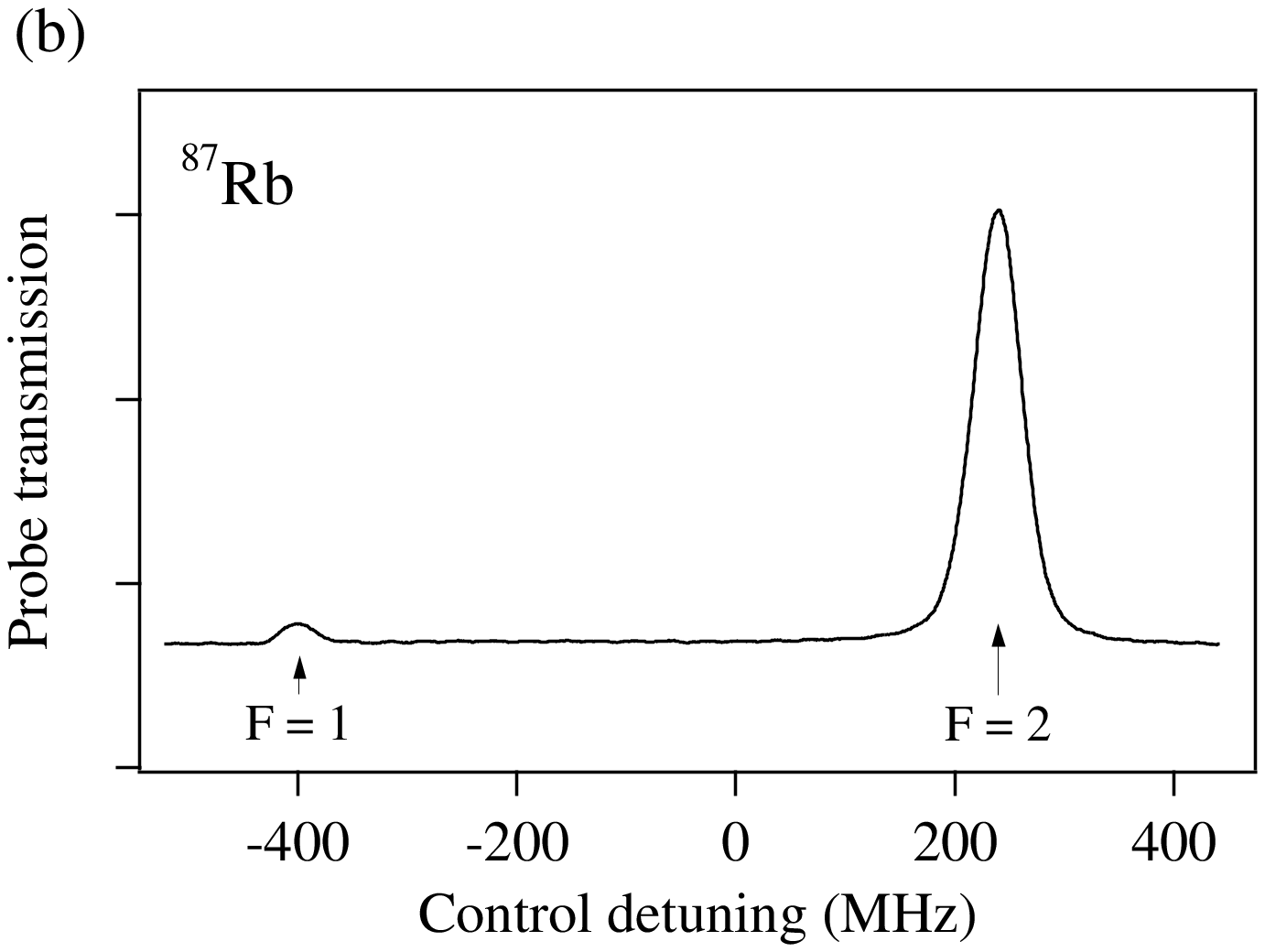}
\caption{Experimental EIT peaks in Rb. In (a), we show the
probe transmission as a function of control detuning in
$^{85}$Rb. The transparency peaks correspond to the two
hyperfine levels labelled by their $F$ values. Control
detuning is measured from the unperturbed $7S_{1/2}$ state.
In (b), we show the corresponding spectrum in $^{87}$Rb.}
\label{f4}
\end{figure}

We now consider experimental spectra in $^{85}$Rb. The
$7S_{1/2}$ state has two hyperfine levels: $F=2$ and 3 with
an interval of 282 MHz. The probe laser is locked to the
$F=3 \rightarrow 3$ hyperfine peak of the lower transition
and has a power of 2 mW. The measured probe transmission
spectrum is shown in Fig.\ \ref{f4}(a). As the control
laser is scanned, there are two distinct transparency peaks
corresponding to the two hyperfine levels. The linewidth of
the peaks is about 50 MHz, which is the theoretically
predicted linewidth at room temperature for a Rabi
frequency of 100 MHz. Note that the corresponding linewidth
at $T=0$ would be about 1500 MHz. The value of
$\Omega_c=100$ MHz is consistent with the control power of
100 mW used for this experiment. A similar spectrum for
$^{87}$Rb is shown in Fig.\ \ref{f4}(b), measured with the
probe laser locked to the $F=2 \rightarrow 2$ hyperfine
peak. Note that the two hyperfine levels ($F=1,2$) in this
case have a larger separation of 638 MHz.

\begin{figure}
\twoimages[width=7cm]{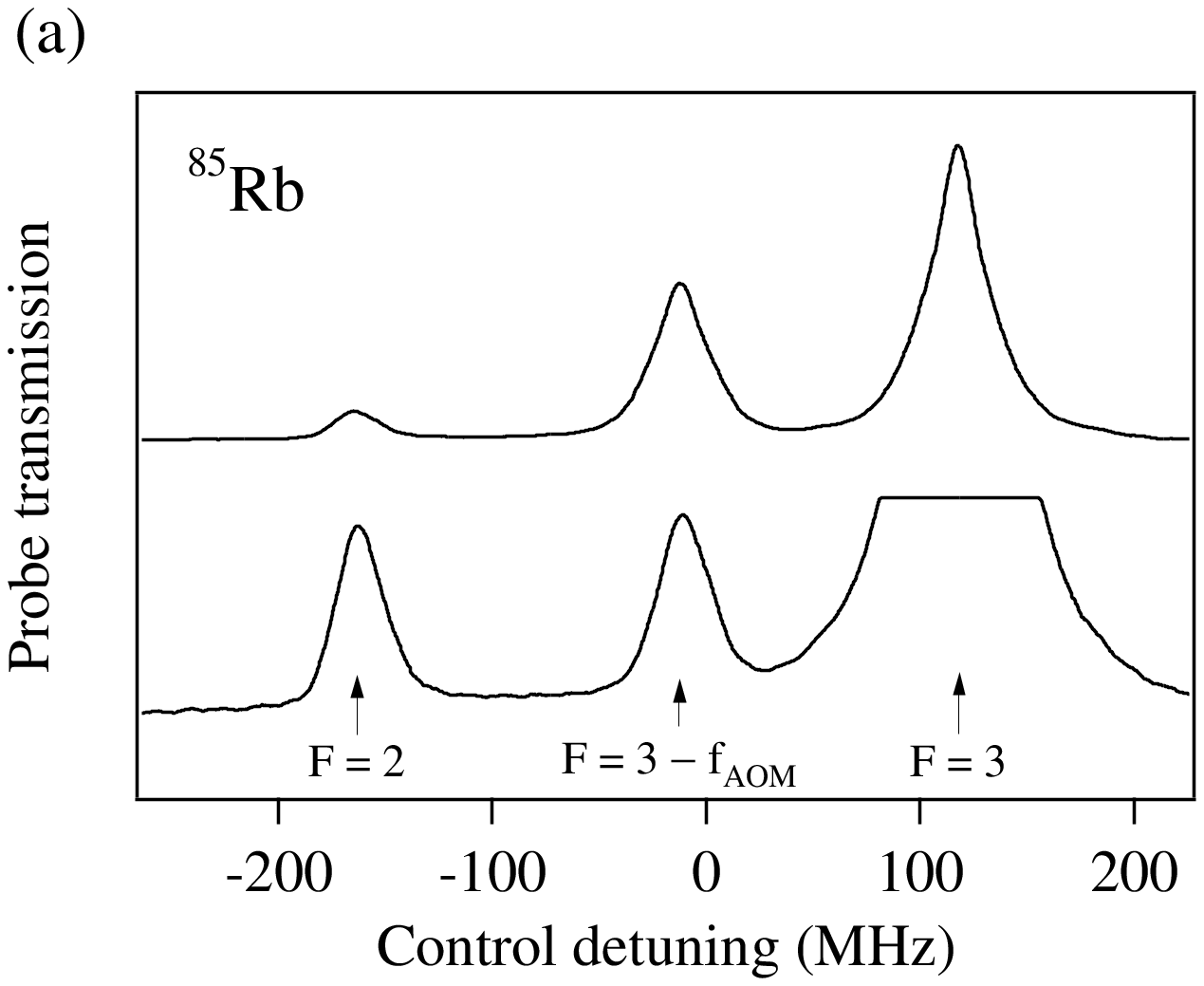}{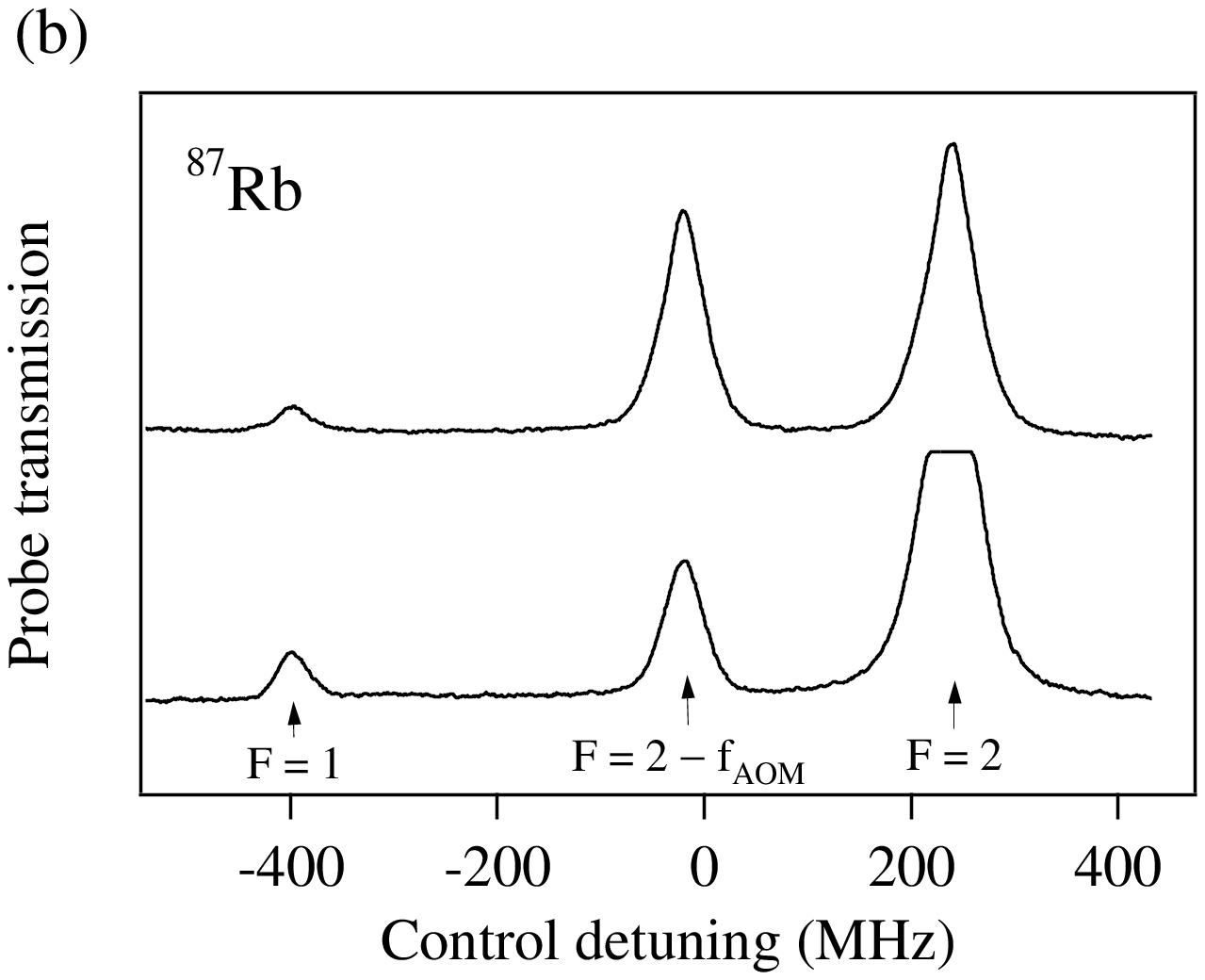}
\caption{Hyperfine measurement using AOM shifting. In (a),
the upper trace is the probe transmission in $^{85}$Rb in
the presence of the AOM-shifted control beam. The new peak
in the middle is the frequency-shifted twin of the $F=3$
peak, allowing precise calibration of the scan axis. The
bottom trace is obtained by reducing the rf power into the
AOM and increasing the photodiode gain so that the smaller
peaks are comparable. In (b), we show a similar set of
curves for $^{87}$Rb.} \label{f5}
\end{figure}

For the measurement of the hyperfine interval, there are
two points to note from the experimental curves. The first
is that we need a precise calibration of the frequency-scan
axis of the control laser. The second is that the height of
the two peaks is very different. Therefore, the measurement
proceeds as follows. The scan calibration is achieved by
using the AOM-shifted control beam, as shown in the
experimental schematic (Fig.\ \ref{f2}). In the presence of
both control beams, the larger peak shows a twin peak at a
location down-shifted by the AOM frequency. The resulting
spectrum of three peaks for $^{85}$Rb is shown in the upper
trace of Fig.\ \ref{f5}(a). Note that the smaller peak also
has a twin which will appear to its left, but this is very
small and not shown. Since the distance between the primary
peak and its twin is exactly the AOM frequency, this allows
us to calibrate the frequency axis precisely. Now, the rf
power into the AOM is reduced so that the AOM-shifted peak
becomes comparable in height to the smaller peak, and
simultaneously the gain of the photodiode amplifier is
increased. The resulting spectrum is shown in the lower
trace Fig.\ \ref{f5}(a). The calibrated distance between
the smaller peak and the AOM-shifted peak is then used to
determine the hyperfine interval. Note that the primary
peak saturates the amplifier, but this is not important
since it is not used in the hyperfine measurement. In other
words, the two large peaks in the upper trace are used for
calibration, while the two smaller peaks in the lower trace
are used for the measurement. A similar set of curves for
$^{87}$Rb is shown in Fig.\ \ref{f5}(b).

The primary advantage of this procedure is that the error
in determining the peak position is reduced. To the extent
that the two peaks are similar, any systematic error in
determining the peak centers will cancel when taking the
difference. The peak centers are determined in two ways:
first by a peak-fitting routine, and then by determining
the zero crossing of the numerical third derivative. This
gives the peak position with a typical error of about 10
kHz.

Before we turn to the results, let us consider the sources
of error in the technique. An important consideration is a
systematic shift of the probe laser from resonance due to
effects such as stray magnetic fields in the vicinity of
the vapor cell, phase shifts in the feedback loop, and
background collisions. However, it is important to note
that this will not affect the measurement of the hyperfine
interval at all. In other words, such a shift is equivalent
to a detuning of the probe from resonance, and it can be
shown from Eq.\ 1 that a non-zero value of $\Delta_p$ will
simply shift the location of the transparency peak by this
amount in the opposite direction. Hence, both peaks in
Fig.\ \ref{f3} will shift by the same amount and the
hyperfine interval will be unaffected.

Similarly, the effect of an angle between the probe and
control beams is unimportant. Such an angle implies that
the two beams come into simultaneous resonance with a
non-zero velocity group. Since the velocity can be either
positive or negative, this only causes a small broadening
of the EIT peak but does not affect the peak center.
Indeed, we have seen that there is no measurable change in
the linewidth of the theoretical curves shown in Fig.\
\ref{f3}(b) for a misalignment angle of 10 mrad, which is
the maximum value of the angle between our beams. However,
there could be a systematic shift of line center if there
is velocity redistribution due to radiation pressure
effects, but this should be the same for both peaks and
cancel in a difference measurement. In a similar manner,
shifts due to collisions will affect both peaks equally and
cancel in the difference.

\begin{figure}
\twoimages[width=6.5cm]{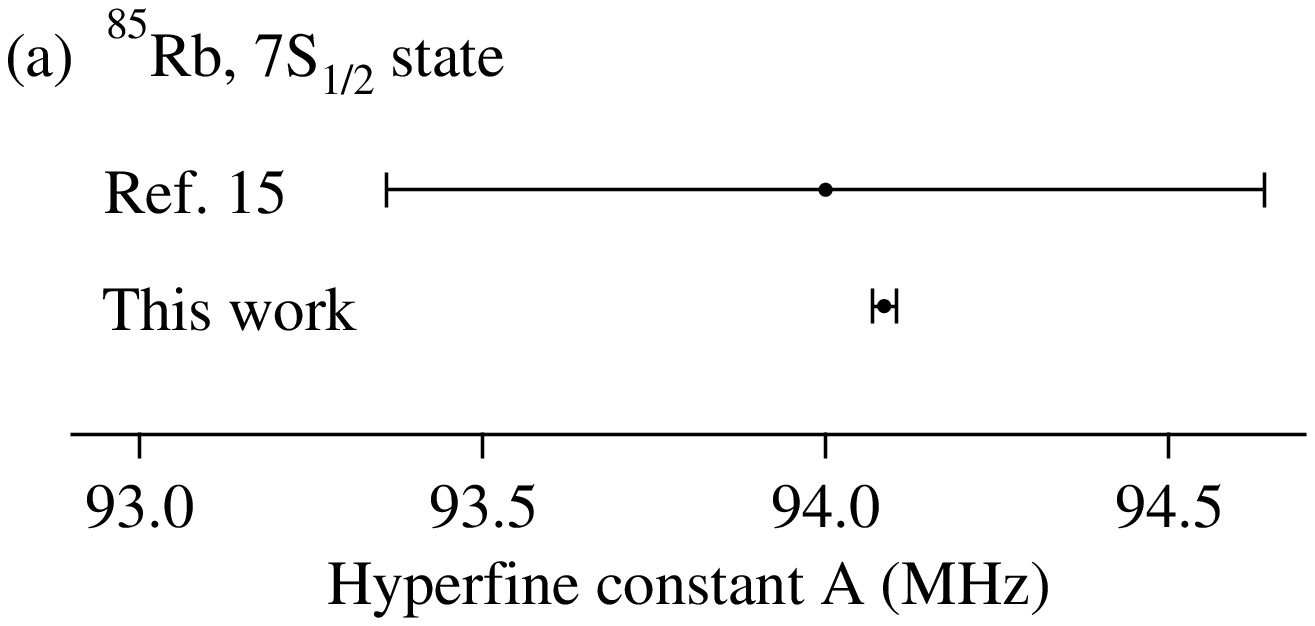}{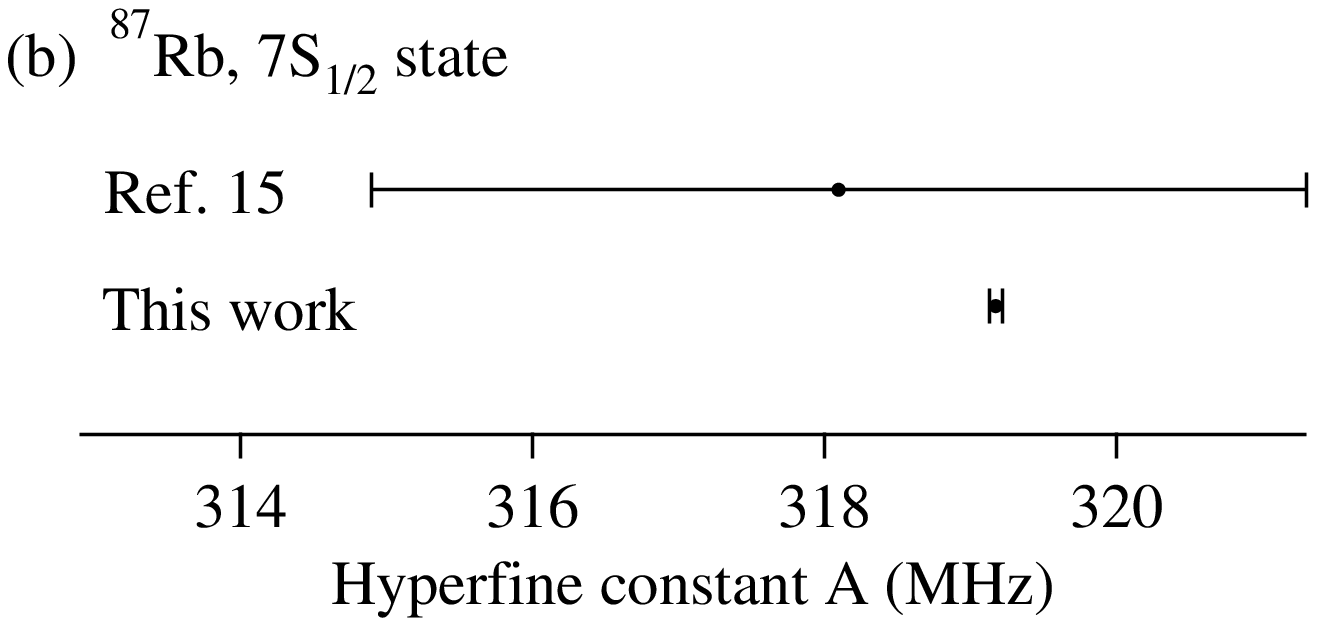}
\caption{Comparison of hyperfine constants. In (a), we
compare the value of $A$ in $^{85}$Rb measured in this work
to the value from Ref.\ \cite{AIV77}. In (b), we show a
similar comparison for $^{87}$Rb.} \label{f6}
\end{figure}

Thus the main sources of error in our technique arise due
to residual nonlinearity of the frequency-scan axis, and
the accuracy in determining the peak centers. To check for
these errors, we have repeated the measurements at various
values of the AOM frequency. A representative list of 6
measurements to show the range of AOM frequencies is given
in Table \ref{t1}. The final values are obtained by taking
an average of all the measurements and calculating the
error in the mean. From a total of about 60 measurements in
$^{85}$Rb and 40 measurements in $^{87}$Rb repeated over a
period of several days, we obtain the following average
values for the hyperfine interval:
\begin{tabbing}
\hspace{1.5cm} \= $^{85}$Rb: $\{F=3-F=2\}$ = 282.254(54) MHz, \\
               \> $^{87}$Rb: $\{F=2-F=1\}$ = 638.347(90) MHz.
\end{tabbing}
The quoted errors include a systematic error of 40 kHz to
account for unknown sources of error. The larger error in
$^{87}$Rb is because of the larger interval being measured,
requiring larger extrapolation of the scan axis. The
measured interval can be used to calculate the value of the
magnetic-dipole coupling constant $A$ in the $7S_{1/2}$
state as:
\begin{tabbing}
\hspace{1.5cm} \= $^{85}$Rb: $A=94.085(18)$ MHz, \\
               \> $^{87}$Rb: $A=319.174(45)$ MHz.
\end{tabbing}
These values are compared with the recommended values from
Ref.\ \cite{AIV77} in Fig.\ \ref{f6}. As can be seen, the
values are consistent, but the accuracy is improved
considerably, by a factor of 35 in $^{85}$Rb and by a
factor of 71 in $^{87}$Rb.

\begin{table}
\caption{Hyperfine-interval measurements using different
AOM offsets in $^{85}$Rb and $^{87}$Rb.} \label{t1}
\begin{center}
\begin{largetabular}{|ccc|ccc|}
\hline \hline
Hyperfine interval  & AOM offset & Value & Hyperfine interval & AOM offset & Value \\
$^{85}$Rb           & (MHz)      & (MHz) & $^{87}$Rb          & (MHz)      & (MHz) \\
\hline
           & 130.021 & 282.280 &            & 180.015 & 638.213 \\
           & 142.001 & 282.210 &            & 190.046 & 638.266 \\
$F=3-F=2$  & 154.012 & 282.277 & $F=2-F=1$  & 200.076 & 638.338 \\
           & 166.005 & 282.224 &            & 240.145 & 638.330 \\
           & 178.093 & 282.297 &            & 250.106 & 638.303 \\
           & 190.017 & 282.291 &            & 270.198 & 638.578 \\
\hline \hline
\end{largetabular}
\end{center}
\end{table}

In summary, we have demonstrated a new technique for
high-resolution hyperfine measurements in excited states
using the phenomenon of electromagnetically-induced
transparency. The experiments have been done using a ladder
system in atomic Rb. The weak probe laser is locked to the
lower transition while the strong control laser is scanned
across the upper transition. Transparency peaks appear in
the probe transmission whenever the control laser comes
into resonance with a hyperfine level. The frequency axis
of the control laser is set by using an AOM-shifted beam
with a known frequency offset. In this manner, we are able
to measure hyperfine intervals in the $7S_{1/2}$ state of
Rb with an error of less than 100 kHz. This is already a
considerable improvement over previous measurements, but we
have not really pushed the limits of accuracy of this
technique. For example, we have recently demonstrated a
technique for measurement of hyperfine intervals with 20
kHz accuracy using an AOM whose frequency is locked to the
hyperfine interval of interest \cite{RKN03}. This will
eliminate any potential errors due to nonlinearity of the
scan. Using this method, we hope to be able to further
improve the accuracy of the current technique. We plan to
measure excited-state hyperfine structure in other atoms
that are of interest for PNC measurements, such as Cs and
Yb \cite{NAT05}.

\acknowledgments

We thank Anand Ramanathan for help with the experiments.
This work was supported by the Department of Science and
Technology, Government of India. One of us (A.W.)
acknowledges financial support from CSIR, India.


\end{document}